\documentclass[aoas]{imsart}

\RequirePackage{amsthm,amsmath,amsfonts,amssymb}
\RequirePackage[authoryear]{natbib}
\RequirePackage[colorlinks,citecolor=blue,urlcolor=blue]{hyperref}
\RequirePackage{graphicx}

\startlocaldefs

\endlocaldefs

\begin{document}

\begin{frontmatter}
\title{A statistical machine learning approach for benchmarking in the presence of complex contextual factors and peer groups}
\runtitle{Statistical machine learning for benchmarking}

\begin{aug}
\author[A]{\fnms{Daniel W.} \snm{Kennedy}\ead[label=e1]{dan.w.kennedy@outlook.com}},
\author[A]{\fnms{Jessica} \snm{Cameron}\ead[label=e2]{???@???}}
\author[A]{\fnms{Paul P.-Y.} \snm{Wu}\ead[label=e3]{p8.wu@qut.edu.au}}
\and
\author[A]{\fnms{Kerrie} \snm{Mengersen}\ead[label=e4]{k.mengersen@qut.edu.au}}
\address[A]{Science and Engineering Faculty, Queensland University of Technology, Brisbane, Australia}

\end{aug}

\begin{abstract}
The ability to compare between individuals or organisations fairly is important for the development of robust and meaningful quantitative benchmarks.
To make fair comparisons, contextual factors must be taken into account, and comparisons should only be made between similar organisations such as peer groups.
Previous benchmarking methods have used linear regression to adjust for contextual factors, however linear regression is known to be sub-optimal when nonlinear relationships exist between the comparative measure and covariates.
In this paper we propose a random forest model for benchmarking that can adjust for these potential nonlinear relationships, and validate the approach in a case-study of high noise data.
We provide new visualisations and numerical summaries of the fitted models and comparative measures to facilitate interpretation by both analysts and non-technical audiences. 
Comparisons can be made across the cohort or within peer groups, and bootstrapping provides a means of estimating uncertainty in both adjusted measures and rankings.
We conclude that random forest models can facilitate fair comparisons between organisations for quantitative measures including in cases on complex contextual factor relationships, and that the models and outputs are readily interpreted by stakeholders.
\end{abstract}


\end{frontmatter}


\section{Introduction}

Benchmarking has been used to make quantitative comparisons in many sectors, such as health, government and education~\citep{Klassen2009}.
Two challenges with benchmarking are the variability and uncertainty of contextual factors hampering direct comparisons~\citep{Dimick2010}, and the correct interpretation of results by the intended audience~\citep{Wilson2008,Pope2009}.
Quantitative measures must therefore be adjusted for contextual factors before a fair comparison to be made between organisations.
For example, when evaluating overall academic performance in schools, a contextual factor is the composition of backgrounds of students, which is known to affect academic performance~\citep{Redden2012}.
Any difference in raw academic performance between two schools may be the result of real differences in the effectiveness of teaching strategies, or could be the result of different student backgrounds, so the effect of background must be taken into account before a fair comparison can be made.

Generally, organisations should only be compared with other similar organisations, because even if efforts are made to adjust for differences in contextual factors, comparison between dissimilar organisations risks substantial influence from unknown (and thus not adjusted-for) factors.
In addition, comparisons become less valid as the qualitative differences between organisations grow.
This is solved by separating groups into similar organisations, called here peer groups, within which comparisons are safer.
A major early finding for strategic groups, a related theoretical group for commercial firms, was that similar strategies of firms were associated with differential performance outcomes depending on the strategic group~\citep{Cool1987}. Thus the organisation's context is important for interpreting performance, and that comparisons may need to be restricted to within groups of similar organisations to ensure fairness.

Quantitative assessment can provide a means of ranking and/or partial comparison, though appropriate adjustment for the known contextual factors is required for it to be fair and useful.
Most previous methods have used a linear regression approach to adjust for such factors~\citep{OMalley2005,Kreutzer2007,Holzer2011}.
In one study~\citep{Holzer2011}, the authors noted that the residuals are skewed to the right, which indicates the linear model is a poor fit, since residuals are assumed to be symmetric in linear regression.
This could be a sign that the model was under-fitting, and that more complex relationships were present, or that the distribution of unmodelled error was asymmetric, rendering a  linear model inappropriate depending on sample size.
In another study, subtle departures from linearity were detected by binning the continuous variable and creating a set of dummy variables~\citep{OMalley2005}.

Another methodology is to first partition organisations using the contextual factors by clustering on similarity into a set of clusters of relatively homogeneous individuals~\citep{ZafraGmez2009}, which may comprise the peer groups.
The raw measurement of two organisations may then be compared without the need to adjust, since substantial differences in contextual factors have been accounted for in the clustering.

There are some examples of more complex regression models for benchmarking. \citet{Zimmerman2006} used cubic splines to model the effects for three continuous variables. The method predicted the length of stay of patients in Intensive Care Units, adjusting these values for continuous contextual factors such as age, acute physiology score and prior length of stay such that their effect was nonlinear. A subset of the data was used as a hold-out to measure the predictive accuracy of the fitted model. While it is clear that a linear model would not have captured clear nonlinear relationships in the data, it is not clear to what extent the use of splines improved the predictive accuracy.

The task of adjusting a quantitative measure so comparisons can be made across different values of one or more contextual factors fits into the broader field of risk adjustment, which is popular in both actuarial science~\citep{Li2013} and health \citep{Warmerdam2018}.
Several regression tree and random forest approaches have been developed to adjust risk factors in the presence of large numbers of contextual factors \citep{Li2013} and interactive effects \citep{Buchner2015,vanVeen2017}.
The importance of interactions in adjustment was recognised by \citet{Goldstein1996a} in a study of student examination results.
\citet{Goldstein1996} noted that the inclusion or exclusion of a variable in an adjustment model substantially influenced the resultant adjusted values.
The authors emphasised the importance of including all possible influences beyond control of the organisation in the model, or the resulting measures will be misleading.

In this paper, we propose using a random forest model with a statistical learning approach to adjust quantitative measures for contextual factors accurately in the presence of both nonlinearity adjusting for peer-groups as well, and compare its accuracy to a linear regression model.
We estimate the uncertainty in adjusted comparative measures and rankings, and use the approximate predictive distribution as a benchmark against which an individual organisation's adjusted measure can be compared and understood.
To ensure all stakeholders can understand the random forest adjustment model and outputs, we 
The resulting statistical learning model provides a better representation of the underlying value addition of organisations and a better basis for comparison, while also retaining interpret-ability for all stakeholders.

\section{Methods}

To provide meaningful and fair comparisons based on the quantitative measure, we introduce the benchmarking problem as being equivalent to the statistical learning problem of regression.
We use the random forest, an algorithmic model popular in statistical learning, as the means of adjusting the raw quantitative measure for the contextual factors.
In order to estimate uncertainty in benchmarking outputs (adjusted measures and rankings) we use bootstrapping, which is also a key component in the random forest model.
We then use an approximate predictive distribution as a benchmark for which individual measures can be compared.
Finally, we introduce two visualisations used to communicate the benchmarking outputs without the need for domain knowledge in statistical learning.

\subsection{Measure Adjustment as Regression}

The problem of adjusting comparative measures for contextual factors corresponds to the statistical machine learning problem of regression, which seeks to find the most accurate model for predicting a response, which corresponds to the raw quantitative measure, given covariates, which correspond to the contextual factors.
Therefore, throughout this work these two pairs of terms will be used interchangeably.
Once they have been taken into account, the residual variation has no dependence on the contextual factors, so it provides a better representation of differences between organisations.

Mathematically, the adjusted comparative measures corresponds to the regression residual.
Let there be $n$ organisations in the cohort.
Let the observed comparative measure value, which will be referred to as the raw measure, for unit $i$ be denoted $y_i$, with all $n$ values denoted by $\mathbf{y} = \left(y_1,...,y_n\right)$.
Let there be $J$ contextual factors, which are coded into the vector $\mathbf{x}_i \in \mathbb{R}^L$ where $J \leq K$, with the entire cohorts contextual factors given by $\mathbf{X}=\left(\mathbf{x}_1,...,\mathbf{x}_n\right)$.
Then we may propose that the variation in $y_i$ is partially attributable to the contextual factors $\mathbf{x}_i$, with this relationship described by a function $f{\left(\mathbf{x}\right)}$.
Therefore, we can write an equation for $y_i$ as 
\begin{equation}
	y_i = f{\left(\mathbf{x}_i\right)} +  e_i
\end{equation}
where the residual $e_i$ represents random variation not attributable to the contextual factors, and is distributed by an error distribution.
The function $f$ represents the expected raw measure value given the contextual factors $\mathbf{x}_i$.
A correctly adjusted measure should allow different units $i$ and $j$ to be compared, even if the values of their contextual factors are different. 
Given an approximation of $f$, $\hat{f}$, the residual is calculated to be the difference between the observed value $y_i$ and the estimated value given the contextual factors $\hat{f}{\left(x_i\right)}$:
\begin{equation}
	\hat{e}_i = y_i - \hat{f}{\left(\mathbf{x}_i\right)}
\end{equation}
The approximation to the function $\hat{f}$ is done by fitting a regression model to the available data. Here we use a random forest model, which is described in Section~\ref{sec:random-forest}.

\subsection{Bootstrap Uncertainty Estimation}

Given the residuals are only estimates of the true deviation of the observed response $y_i$ from the expected value $f{\left(\mathbf{x}_i\right)}$ given contextual factors, it is important to quantify the uncertainty around the point estimates.
This uncertainty is then used to evaluate whether there is evidence for differences in the adjusted measure in pairwise comparisons, as well as in ranking with uncertainty.
Bootstrapping~\citep{Efron1979} is a method of approximating the error sampling distribution using only the data itself, without the need for a parametric model.
Bootstrap resamples are created by first sampling observations from the data $n$ times with replacement to form a resample $D^{\left(b\right)} = \left(\mathbf{y}^{\left(b\right)},\mathbf{X}^{\left(b\right)}\right)$. 
The regression model is then fit to $D^{\left(b\right)}$ to get an approximate $\hat{f}^{\left(b\right)}$.
The statistics of interest are the residuals $\hat{e}^{\left(b\right)}_i = y_i - \hat{f}^{\left(b\right)}{\left(\mathbf{x}_i\right)}$, which are called the bootstrap replicates.
When this is done repetitively many times, an approximate sampling distribution of the residuals is formed from the distribution of the replicates.

\subsection{Random Forests}\label{sec:random-forest}

The random forest~\citep{Breiman2001} is an ensemble learning approach, and is composed of the average prediction of many individual decision tree models.
Each tree is `grown' from a random bootstrap resample of the data. At each node in the tree, a random subsample of the covariates is drawn, and then from this subsample the best cut is chosen to minimise prediction error. 
The prediction given by the forest is equal to the average prediction from the trees.

The number of covariates randomly drawn per node is a tuning parameter $m_{try}$, which is chosen by cross-validation to prevent over- or under-fitting.
Two popular measures of generalised accuracy are the root mean squared error of prediction, and the $R^2$, which estimates the proportion of variance in the response explained by the model.
The other parameter, $n_{tree}$, is the number of trees in the forest, and it does not require tuning since there is no risk of over-fitting.
The computation time of the fitting algorithm does however increase with the number of trees and the improvement in accuracy becomes progressively smaller as new trees are added.
It is therefore common practice to inspect the prediction accuracy of subsets of trees of different sizes to investigate if the prediction accuracy has levelled out or if more trees need to be added.

An advantage of the random forest over other models is that it has an inbuilt estimate of generalised (also called out-of-sample) error, called the out-of-bag error.
Usually cross-validation is required, separating the data the model is fitted on to the data where error is estimated, however this is not required in random forests.
For a given observation in the data, by selecting only trees in the forest which are not fitted using it, one can use their averaged prediction instead of the average over the entire forest.
Provided there are sufficient trees in the forest, every observation will have some trees which are not fitted using it, and so the process can be used to get an out-of-bag prediction for every observation.
The resulting error estimate calculated from out-of-bag predictions is an unbiased estimate of the generalised error.

\subsection{Ranking}

Ranking is the problem of ordering the organisations from highest to lowest based on adjusted measure.
A single ``point estimate'' of rank can be extracted from the residual estimates, but like the residuals themselves the ranking is not known exactly.
Bootstrapping allows for ranking uncertainty to be quantified by ranking each set bootstrap residual replicates for many bootstrap resamples, providing a sampling distribution for the rank of each observation.

\subsection{Approximate predictive distribution}

To evaluate adjusted comparative measures, one needs to compare it against a reasonable benchmark distribution, which characterises the typical range one would expect for the organisation.
For this we chose to use the approximate predictive distribution for the raw measure for a ``new'' organisation with the same associated contextual factors as the given organisation.

Bootstrapping can be used to approximate the predictive distribution by refitting the model for many bootstrap resamples and calculating bootstrap residual replicates for each observation. This has been shown for linear regression~\citet{Stine1985} and non-parametric regression~\citet{Kumar2012}.
Combining the replicates of all observations gives the approximate predictive distribution.
However, even if the out-of-bag residuals are used in random forest, because an observation is sometimes chosen more than once for a bootstrap resample, the out-of-bag residual may use one bootstrap draw of an observation to predict for another bootstrap draw from the same observation.
One could find the subset of trees which contain none of the bootstrap draws from an observation, but in practice this can result in only a few trees being available if the observation was drawn multiple times for a bootstrap resample.

Instead, we considered a modification to approximate the predictive distribution using only the residuals of observations not in the current bootstrap resample, called here the Out-of-Resample residuals. To estimate this distribution we used a bootstrapping scheme:

For $B$ resamples $b = 1,...,B$:
\begin{enumerate}
	\item Resample data with replacement to form $D^{(b)}$, with $D_{-}^{(b)}$ being the observations not in $D^{(b)}$.
	\item Fit a random forest on $D^{(b)}$, $\hat{f}^{(b)}$.
	\item For samples in $D_{-}^{(b)}$, compute the residual.
\end{enumerate}

Given there is an approximate 63.2\% chance that an observation will be chosen for given resample~\citep{Efron1997}, only $0.378B$ bootstrap replicates will be retained for each observation on average.
Therefore, more bootstrap iterations are required for the individual observations.
However, to get the predictive distribution, we can use the entire set of replicates from all observations.
A draw from the predictive distribution for a given $\mathbf{x}'$ is equivalent to randomly picking a residual from the error distribution and adding it to the expected value $f{\left(\mathbf{x}\right)}$.
The bootstrap predictive distribution simulates this as randomly picking an observation $i$ from $1$ to $n$, then randomly picking a resample $b$ where $D^{(b)}$ does not contain $i$, and then adding the residual $\hat{e}^{(b)}_i$ to the Out-of-Bag prediction $\hat{f}^{(b)}{\left(\mathbf{x}'\right)}$.
To compare the predictive distribution to an organisation's adjusted measure, the distribution is then centred to 0, which we called the residual sampling distribution.

By combining the bootstrap residual replicates from all organisations, there are approximately $0.378Bn$ samples of this predictive distribution. This method assumes that the residual distribution is homogeneous over the covariate space, because the observation $i$ is randomly sampled from the entire cohort.

A problem with this approach is that because at any given bootstrap resample, around a third of residuals are missing, it is not possible to compare residual values and extract a full ranking.
To allow comparison between all pairs of observations, as is required for ranking, we used a multivariate normal approximation to the bootstrap residual distribution to get rank distributions.
The covariance matrix, computed using pairwise complete observations, was not positive definite, so the closest positive definite matrix was obtained by eigendecomposition.
Means, medians, standard deviations, and interquartile ranges of samples from the approximate normal were compared with the same statistics for the bootstrap residuals to determine if the fit was reasonable.
Additionally, the same analysis was performed to compare pairwise statistics such as mean difference and correlations were computed to determine if the second-order relationships were well approximated.

With an approximate distribution for the predictive distribution as a benchmark, an individual organisation's adjusted measure and associated uncertainty can be viewed relative to this benchmark.
We chose to use the Probability Integral Transform~\citep{Angus1994} to transform the organisation's bootstrap replicates into percentiles of the predictive distribution.
Therefore, if an organisation's confidence interval bounds were within the predictive distribution's lower or upper tail, it would be clear that the organisation's adjusted measure was low or high respectively.

In addition to comparison relative to the cohort, comparison relative to an organisation's peer-group can also be computed.
To do this, instead of combining bootstrap residual replicates from the entire cohort, only the replicates from the organisations in the peer-group are used, and the organisation's bootstrap residual replicates are compared to this peer group predictive distribution.
Under this method, the assumption is that the predictive distribution is homogeneous over the peer-group, rather than the entire cohort.

\subsection{Communicating Random Forests and Adjusted Measures with Uncertainty}

The random forest has a measure of variable importance based on the out-of-bag error~\citep{Breiman2001}.
For each variable, the out-of-bag prediction is calculated after permuting the values of the variable, thus removing its capacity to assist in prediction. 
The resulting increase in the mean squared error is a measure of the variable's importance.
We also utilised a measure of variable group importance, since variables could be grouped into subsets based on correlation.
This measure uses the same principle of permuting values, but on a group level to remove the entire variable groups' contribution to prediction~\citep{Gregorutti2015}.

Hierarchical clustering was utilised to group variables based on correlation, and a cut-off of 0.7 was used, such that variables would only be clustered together if their linear correlation was greater than 0.7 or less than -0.7.

Partial dependence plots are visualisation methods for interpreting the marginal effects of predictor variables. For a particular predictor variable, partial dependence plots display the expected predicted value for a given value of the predictor, after marginalising out the effect of the other predictor variables. The resulting plot indicates where the predicted value increases and decreases in response to the predictor, however it should be treated with caution when there is statistical dependence between the predictor variable in question and the other predictor values.

Because stakeholders often prefer simpler, intuitive visualisations rather than statistical graphics, we developed two methods of presenting the required information for interpreting the adjusted comparative measure. A simple, single dimensional ``spirit-level'' plot was developed for communicating the value of the comparative measure of an organisation relative to its cohort or peer-group. 
The axis of the spirit-level are percentiles from 0 to 100\% of the bootstrap-estimated predictive distribution, where 0\% represents the lowest bootstrap replicate, and 100\% the highest replicate.
Guidance lines are given for 10\%, 25\%, 50\%, 75\%, and 90\%, indicating what range the top 10\%, lowest 10\%, and quartiles of raw measures are likely to take, given the predictive distribution for the organisation.
To further facilitate interpretation, the associated quantile values are given above the guidance lines, and these values are shifted by the predicted value of the organisation in question, so these values are specific to the organisation and can be compared with the organisation's actual measured value.

The organisation's actual adjusted measure is not known exactly as it is a function of the fitted value, which is uncertain. To communicate this uncertainty, we depict it using a horizontal 90\% confidence interval for the organisation's own adjusted measure in terms of the percentiles of the predictive distribution.

To provide an understanding of how the random forest adjusts comparative measures to a non-technical audience, the mean adjusted measure of the most similar organisations is also plotted on the spirit level plot.
In addition, one-dimensional plots for the contextual factors depict the values of the organisation in question, the most similar organisations, and the peer-group. The values are converted to percentiles of the data distribution using probability integral transform for the empirical cumulative distribution. This ensures they are simple to interpret. 
The organisation can thus see if they are close to their neighbours and to their peer-groups for the important variables (as judged by random forest importance).

\subsection{Case Study: Benchmarking of service provision}

The methodology developed was used to apply quantitative benchmarking to a set of organisations involved in providing services to clients.
A total of 14 client demographic summary statistics were calculated for each organisation, which describe socioeconomic status, ethnicity, gender, regionality/remoteness and age.
To preserve confidentiality, these variables have been relabelled as covariate-X, where X is a number, so their exact meaning is obscured.
The raw comparative measure was calculated for each organisation, and this is also anonymised for confidentiality reasons, although it relates to the amount of service provision, where a higher value indicates an organisation has provided services to a greater degree.

Data were transformed where required to reduce skewness: proportions were transformed using a logit distribution, while count and positively skewed variables were log-transformed. The response was log-transformed before the random forest was fitted.
Peer-groups were constructed based on the covariates using a constrained clustering algorithm developed in a previous paper (under review).

All data processing and analysis was carried out in R~\citep{R2020}. All R scripts required to perform the analysis are publicly available online (\url{https://github.com/danwkenn/peer-group-benchmarking}).

\section{Results}

The smallest cross-validation root mean squared error was for an $m_{try}=4$ variables, though the highest $R^2$ was for around $m_{try}=2$ to $m_{try}=3$ (see Appendix).
The differences in accuracy between $m_{try}$ values were however very small compared to the estimated standard deviations for both accuracy metrics, so there was no indication that any given $m_{try}$ value was substantially better than any other .
Therefore, the value of $m_{try}$ was fixed to 4 for the final model.
The cross-validated $R^2$ value was approximately 0.57 (SE: 0.12), meaning the model explains approximately half of the variation in the response with covariate variation.
The parameter $n_{tree}$, which is the number of trees, was chosen to be 300, and the out-of-bag accuracy was found to have reached near-optimal values.

\begin{figure}
	\centering
	\includegraphics[width=0.9\linewidth]{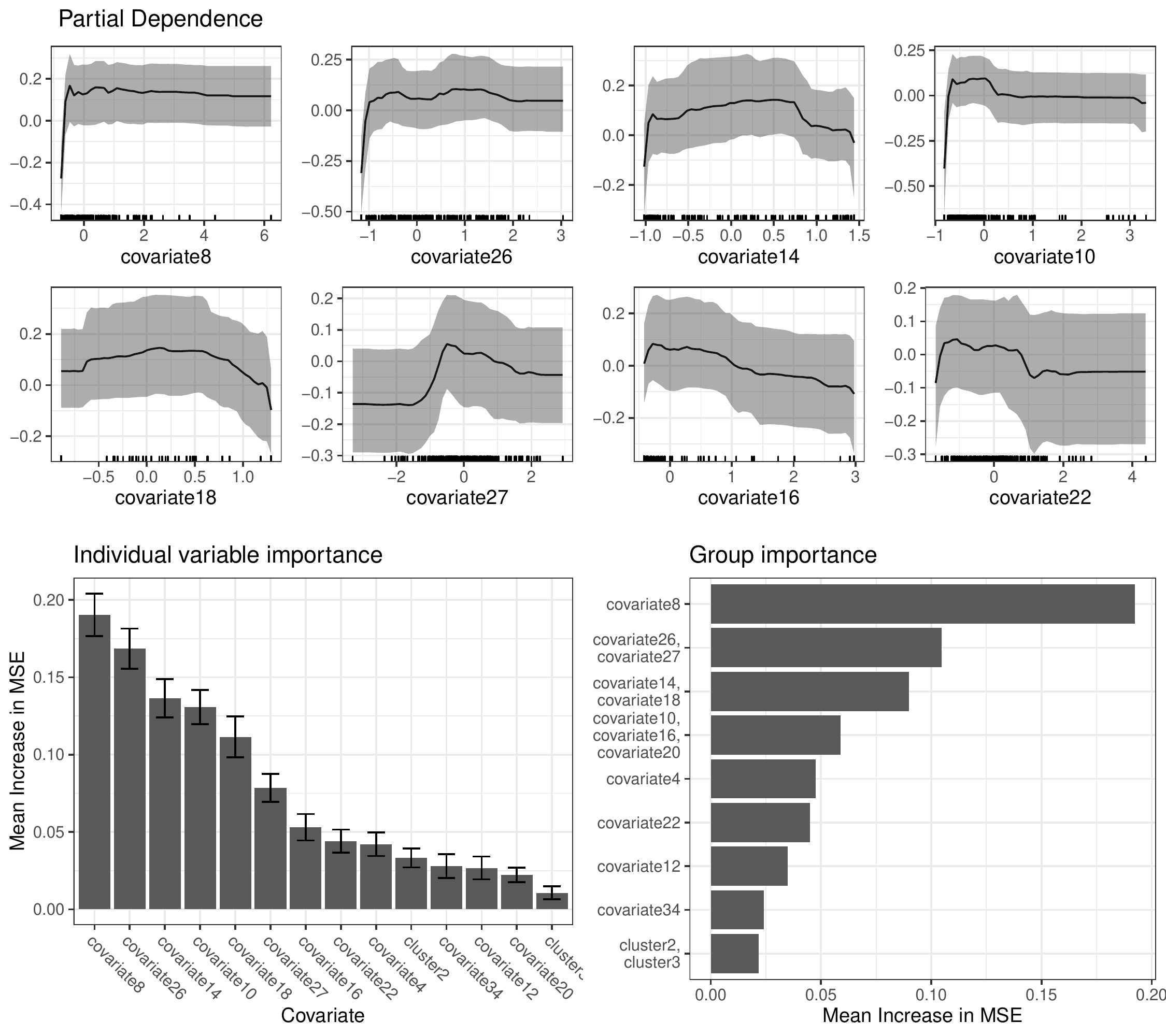}
	\caption{Visualisations for analysing the fitted random forest model. On the top two rows, partial dependence plots showing the effect of predictors after marginalising over the effects of all other predictors. Rug-plots are are included at the bottom of each plot to incidate where the actual data is concentrated. Bottom-left shows variable importance, as measured by increase in mean squared error for all 13 covariates, and peer group. The peer-group differences are represented by the two dummy variables cluster3 and cluster2, representing the change from Peer Group 1 to Peer groups 3 and 2 respectively. Bottom right shows the group variable importance calculated using mean increase in mean squared error. Variables were grouped using heirarchical clustering with a correlation cut-off of 0.7.}
	\label{fig:combined-plot1}
\end{figure}

The partial dependence plots (Figure~\ref{fig:combined-plot1}) showed that the most important variable, covariate 8, had a monotonic positive effect on the response, increasing rapidly before leveling off in a single step.
The variable was also not highly correlated with any other variables, so this is not affected by dependencies.
Other variables, covariate 26 and covariate 10 had similar effects, while covariate 4 had a small, positive, mostly linear effect.
The plot for covariate 14 indicated a non-monotonic effect, however the error bars were wide, likely due to high correlation with other variables.

The variable importance measures (see Figure~\ref{fig:combined-plot1}) indicated two dominant covariates. The most important variable for predicting response was covariate 8, which is related to cultural and linguistic diversity in clients. Covariate 26, which is related to client age, was also important. Interestingly, the peer-grouping dummy variables, labelled cluster2 and cluster3, did not feature prominently. This is to be expected, as the peer-groups were created using the covariate data, so any differences between the peer-groups would likely be accounted for in the covariates.

The group variable importance (see Figure~\ref{fig:combined-plot1}) again noted the importance of covariate 8, as well as covariate 26 grouped with covariate 27, which are both age-related. The group of covariate 14 and covariate 18 were also important, with both of these being related to regionality. There was some indication that a group of covariates 10, 16, and 20, which related to remoteness and indigenous ethnicity were important for predicting the the response. Again, the peer group variables did not feature highly.

\subsection{Comparison with Linear Regression}

The average cross-validation error (root mean squared error) for linear regression was 0.86 (SD: 0.29), compared to 0.69 (SD: 0.13) for the random forest. 
These values are comparable, however the cross-validation error for the two methods are highly correlated and dependent on the data. 
A bootstrap estimate of the difference in out-of-sample error was 0.30 (SD:0.12), showing that the random forest was consistently more accurate than the linear regression.
Based on the $R^2$ values, the random forest explained an average of 30.5\% (SD: 13.1\%) more of the variation in the response than linear regression.

\begin{figure}
	\includegraphics[width=1\linewidth]{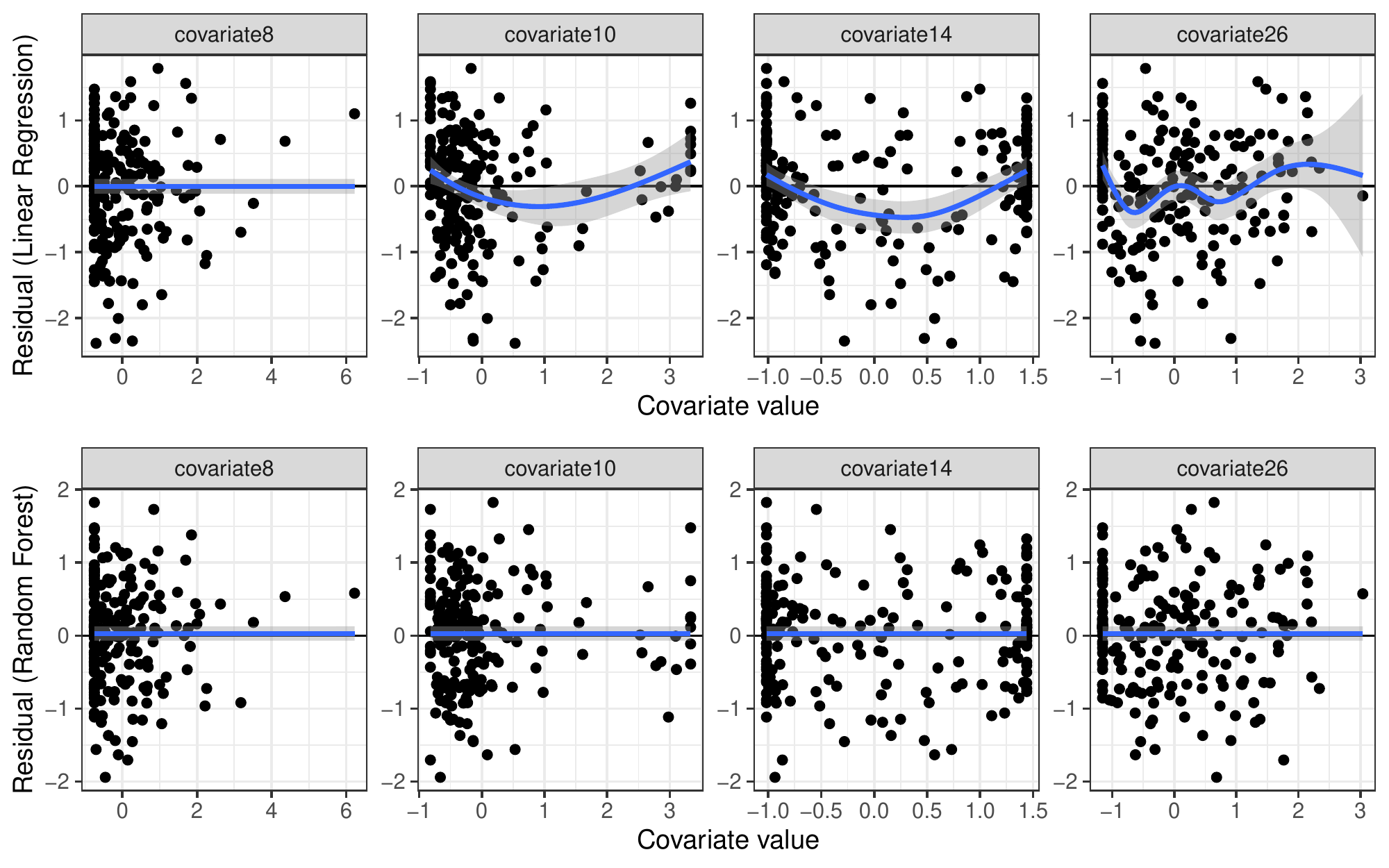}
	\caption{Residual plots for the random forest and linear regression against values of the covariates/contextual factors. A smooth trend-line (blue) is provided with standard error (grey ribbon) to show deviations from the zero-line. Covariates were chosen as representatives of the four most important covariate groups as determined by the group variable importance measures.}
	\label{fig:residual-plots}
\end{figure}

Analysis of the residuals showed that the LR model failed to fully account for the relationship between the response and the covariates 10, 14 and 26, as residual patterns could be seen in the residual scatter-plots (see Figure~\ref{fig:residual-plots}).
This was particularly pronounced for covariate 14, where the residuals of organisations with intermediate values were on average less than 0.
Here, the linear regression model is over-estimating the organisation's expected raw measure, and thus is not flexible enough to account for the nonlinearity in this variable's effect.
The linear regression was also unable to fully account for the step-wise relationship between the raw measure and covariate 8, although this is not visible in the residual plots since the step occurs close to the minimum of the data.
Therefore, the overall result is that the LR model is biased due to under-fitting, resulting in larger residual uncertainty overall, and at an individual organisation level specific biases occur in residual estimation.

\subsection{Estimation of the residual sampling distribution}

\begin{figure}
	\includegraphics[width=1\linewidth]{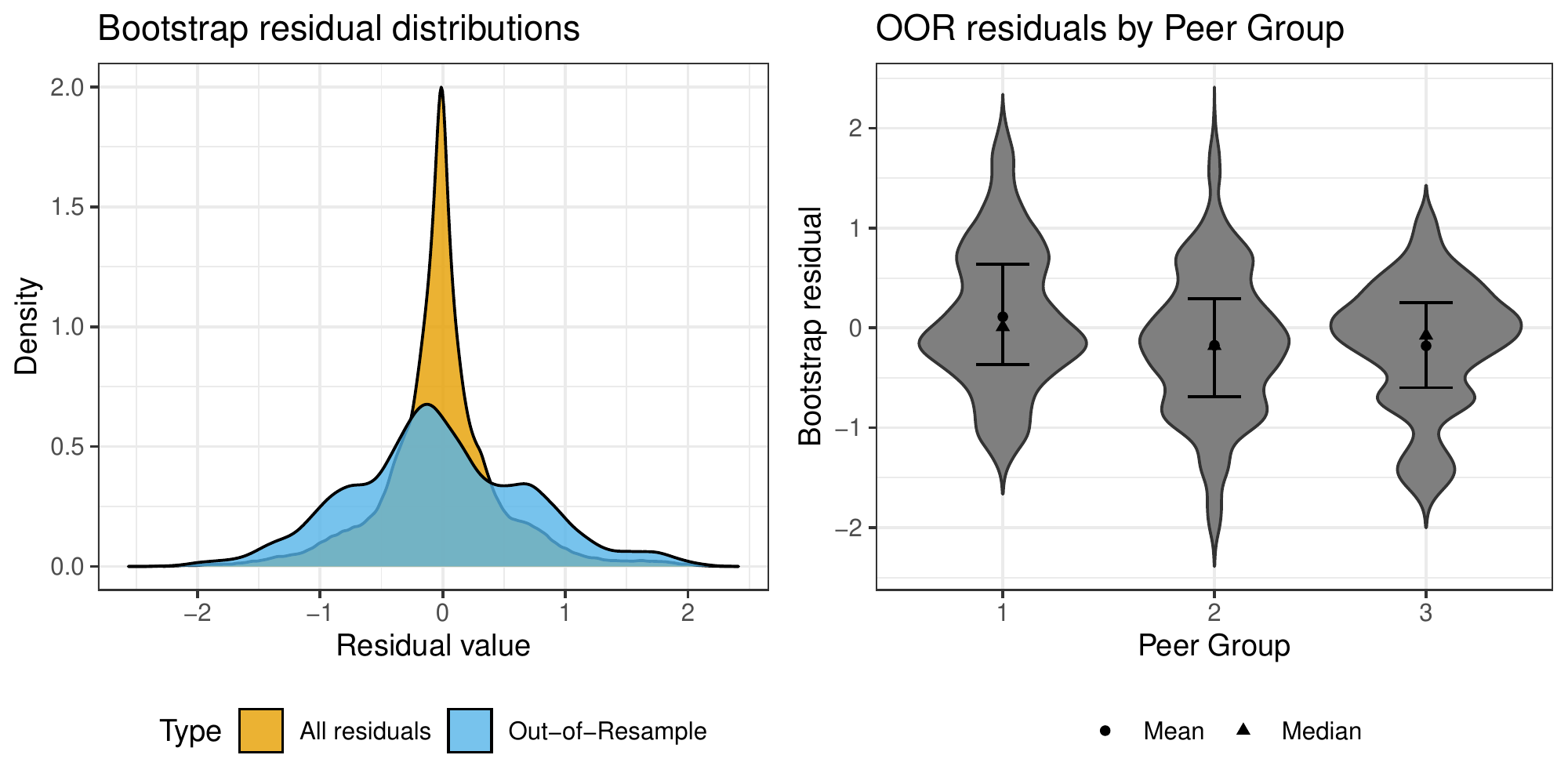}
	\caption{(A) Distributions of different residuals from bootstrapping. The complete residual replicate set contains cases where the data is used to predict itself as a result of being drawn more than once in a bootstrap resample. The out-of-resample residuals are filtered to prevent this. (B) Out-of-resample (OOR) residual replicate distributions for the three peer groups. Interquartile range is depicted by the error bar, and mean and medians are also plotted.}
	\label{fig:bootstrap_plots_combined}
\end{figure}

For 10,000 bootstrap resamples, 200,000 residual replicates were calculated, with approximately 73,000 being out-of-resample residuals.
There was a marked difference in distribution between the entire set of bootstrap residual replicates and the out-of-resample replicates.
We found over 50\% of the bootstrap residual replicates were equal to 0.
In contrast, none of the out-of-resample replicates were equal to 0, which is to be expected, since no observations are used in their own prediction.
The result was the distribution of the entire set having a much lower standard deviation and sharply peaked around 0 (see Figure~\ref{fig:bootstrap_plots_combined}), compared to the subset of out-of-resample replicates.

There was some indication of multimodality in the out-of-resample residual distribution, however the modes did not correspond to peer group.
There were small differences between the width, center and shape of the bootstrap residual distributions for specific peer groups, although they were mostly overlapping.
Peer-group 2 had a wider range than 1 and 3, although this was mostly due to longer tails since the interquartile ranges were comparable.
The similarity of means to medians indicated symmetry of residual distribution in each of the peer-groups, although there was some multimodality present as well.
This implies that with respect to the raw measure, there is approximately the same probability that an organisation will be above its expected raw measure as below. This was not the case for Peer Groups 1 and 3, which exhibited positive and negative skew respectively.

\begin{figure}
	\includegraphics[width=1\linewidth]{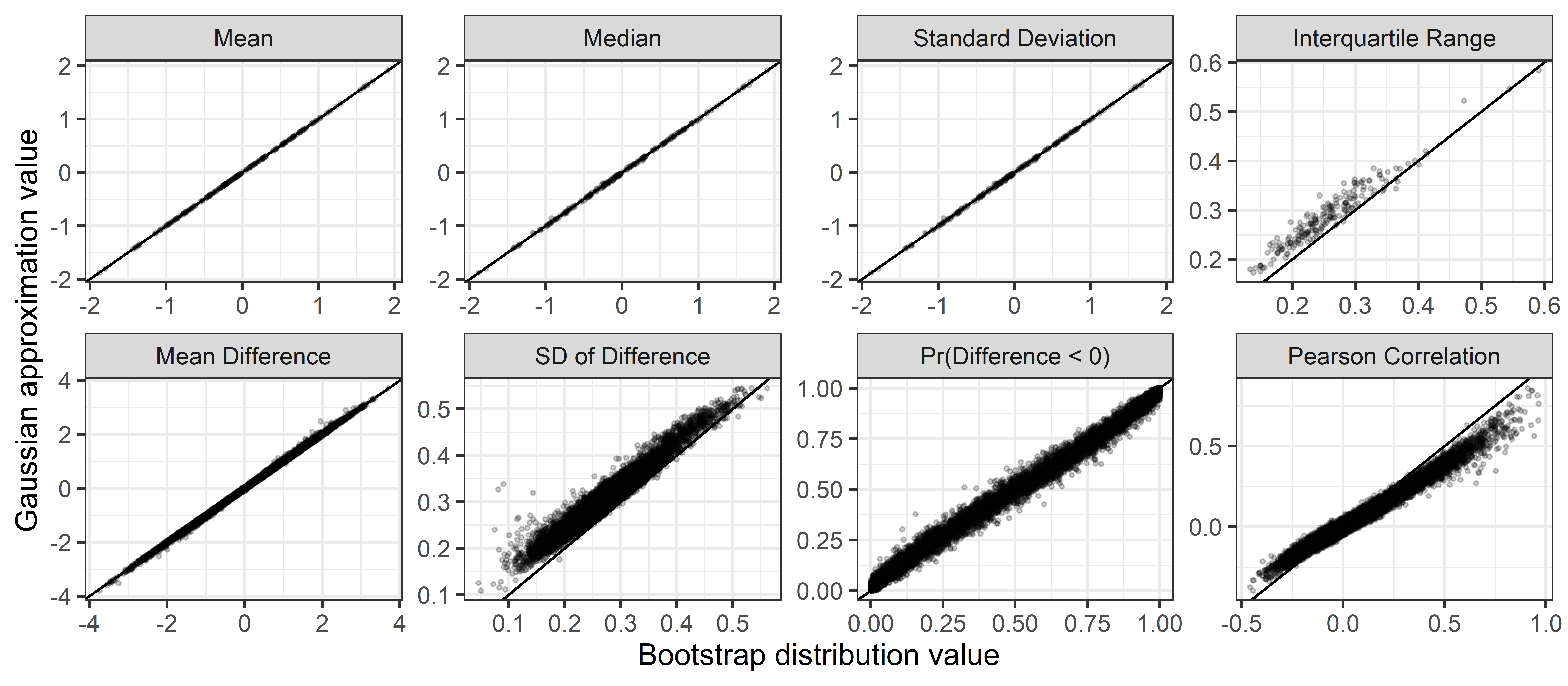}
	\caption{Comparisons of the normal approximation of the bootstrap distribution against the bootstrap distribution itself. On the top row, each plot depicts a different summary statistic, and each point represents an organisation. The bottom row shows comparisons of two-way summary statistics between pairs of organisations. Each point represents a pair of organisations.}
	\label{fig:full_boot_norm_comp_plot}
\end{figure}

We compared several statistics to investigate whether the normal approximation was accurately portraying the centres spreads, and correlations between the bootstrap residuals (see Figure~\ref{fig:full_boot_norm_comp_plot}).
The means of the normal approximation matched the bootstrap residual means exactly, which is to be expected given the method uses the bootstrap means, and the medians were also similarly precise.
The standard deviations were well correlated, but there was a small positive bias of around 0.018. The interquartile ranges were also well correlated with a larger positive bias. 
The pairwise differences were highly accurate, as expected, although there was a small positive bias on the standard deviations of the pairwise differences. There was a good correspondence with the correlations, although for highly correlated pairs (Pearson's $R^2 > 0.5$) the normal approximation counterparts were systematically less correlated.
The probabilities that the difference is less than 0 appeared to be unbiased, with small discrepancies. Therefore, given these probabilities are the basis for ranking we determined that approximation was accurate enough for use in ranking.

\subsection{Rankings}
\begin{figure}
	\centering
	\includegraphics[width=1\linewidth]{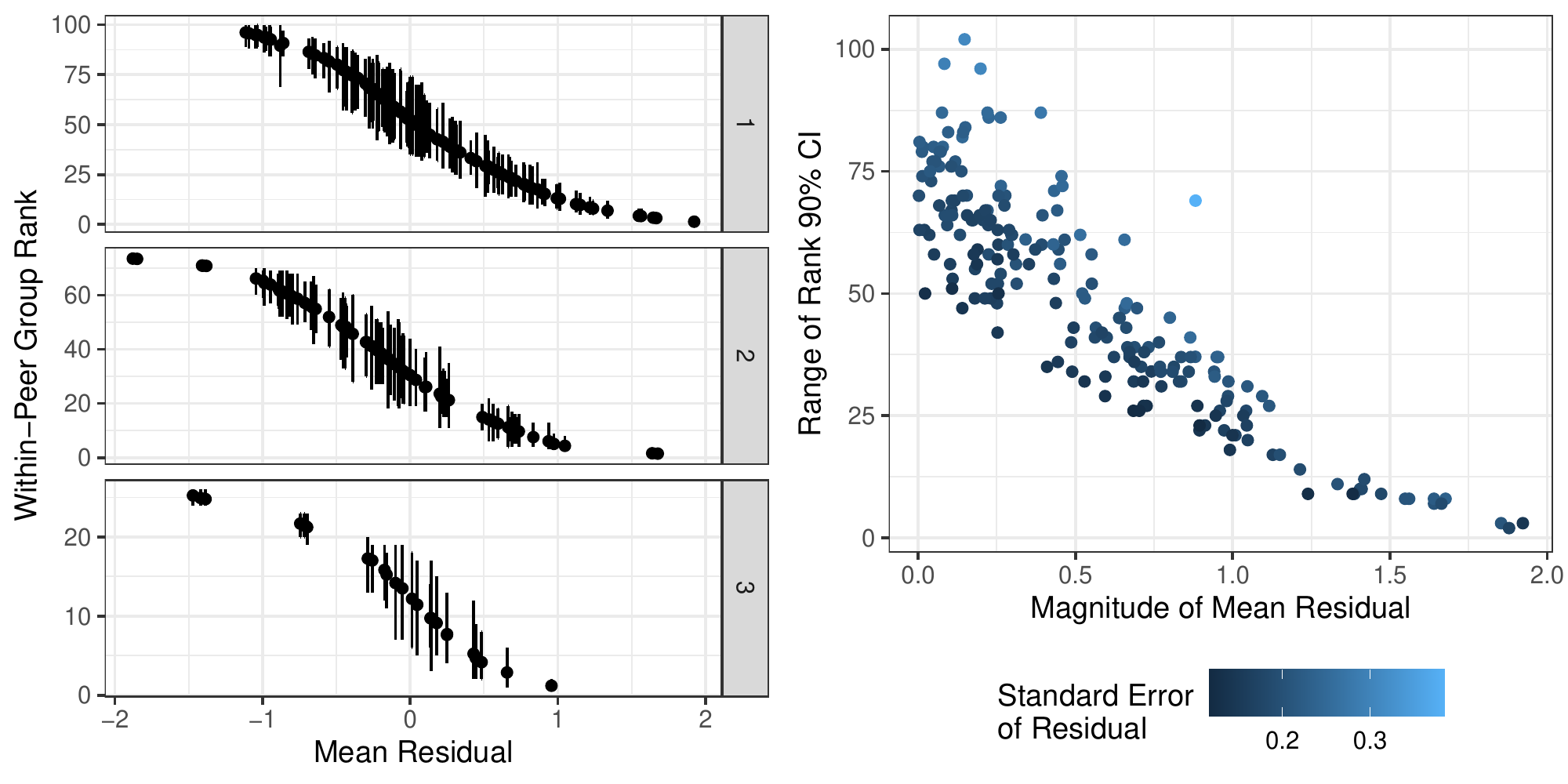}
	\caption{(left) Bootstrap rankings of observations relative to peer group with 90\% confidence intervals. Organisations are separated into the three peer groups. Points indicate modal (most likely) rank of the organisation. X-axis indicates the mean bootstrap OOR residual value of the observation. (right) Range of a 90\% Confidence interval for the rank plotted against the standard error of the residual (colour) and the mean residual (X-axis).}
	\label{fig:rankplotscombined}
\end{figure}

Rankings with associated 90\% confidence intervals, modes (most likely rank) and mean were computed from the normal approximation samples (see Figure~\ref{fig:rankplotscombined}).
There was substantial variation in the width of these intervals, ranging from 2 to 100, with the greatest widths occuring when the mean residual was close to 0.
It was evident that the standard error and mean of residuals was a very good explanation for the size of the confidence interval ranks.
The standard error tended to set an upper limit to the confidence interval range, under which there was some variability, which was controlled by the absolute value of the mean of the residual.
The result of this was a high degree of precision in rank for the highest and lowest-ranked observations and a wide range for middle observations.
Residual standard error was partially controlled by the number of close observations, as there was a small correlation ($R^2$ = 0.11) between the standard error and the mean distance between the nearest 10 observations.
Therefore, the uncertainty in ranking is most acute for non-extreme organisations, with proximity to other organisations being a small factor.

\subsection{Communicating Benchmarking Results}

We devised a two-plot system for communicating the adjusted comparative measure values to non-technical audiences.
The first is the spirit-level plot, which expresses the organisation's measure and associated uncertainty with respect to either the predictive distribution for the entire cohort, or for the individuals associated peer-group.
The residual sampling distribution was recentered to be around the predicted raw measure for the organisation, and guiding values were transformed to the scale of the raw comparative measure.
By using values of the raw comparative measure, an organisations's representative can understand the scale of the plot without having to understand the variable transformations used in the analysis, or the adjustment model used.
In addition, the percentiles of the guiding values are provided, enabling a probabilistic insight into the intervals. Some accompanying text was placed under the spirit-level giving guidance as to the interpretation of the graphic:
\begin{quote}
	The above plot shows your [quantitative measure], adjusted for context (red dot) relative to the entire cohort/peer group. Your measurement also includes a horizontal error bar, because there is some measurement error around the adjustment. The bar indicates the values for which we have 90\% confidence your adjusted value lies, relative to the rest of the cohort. The spirit level shows the range of possible adjusted values for the entire cohort/peer group. Each of the horizontal guidance lines shows a percentiles. For example, the 90\% guidance line on the right of the plot indicates that according to the model, there is a 90\% probability that the value of [quantitative measure] of an organisation with your clients would be equal to or below this value. The other dots on the plot are the adjusted values for the 10 most similar organisations to you, based on the client demographics.
\end{quote}
The densities shown are not useful to a non-technical audience, however they provide a visual link between the spirit-level plots and the bootstrap predictive distributions in which they are based, and may be useful for analysts.
Differences between the cohort- and peer-group spirit plots can be explained by discrepancies between the distributions.

\begin{figure}
	\includegraphics[width=1\linewidth]{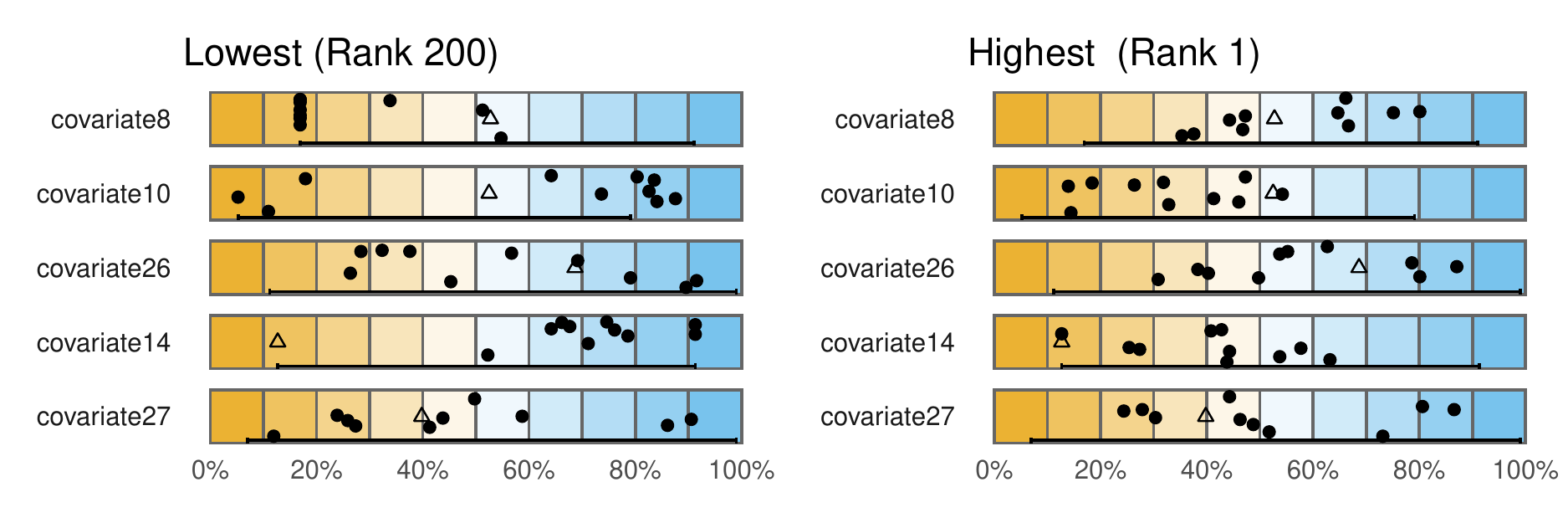}
	\caption{Covariate values depicted accessibly using the percentile plot. The x-axis are percentiles for the given covariate. Points show the given organisation (triangle) and closest 10 neighbours that were also plotted on the spirit plot. Each box represents 10\% of the data. The error bar represents the range of the organisation's peer group.}
	\label{fig:percentile-plots}
\end{figure}

The percentile plots (see Figure~\ref{fig:percentile-plots}) show the organisation's covariate values against the cohort, the peer-group, and the 10 nearest neighbors. The covariates shown are chosen using the random forest variable importance measure. In the case of the highest adjusted measure observation, the organisation is well surrounded by the neighbors and typical the peer group, and this justifies the adjusted measure given in the spirit-level plot. For the lowest adjusted measure organisation however, the value for covariates 8 and 10 were quite different from the 10 nearest neighbors, although not atypical for the peer-group. In such a situation, it may be indicating that in the multidimensional parameter space this organisation is somewhat of an extreme value, and would justify removal from the data-set or making the organisation in question not subject to quantitative comparisons.

\begin{figure}
	\centering
	\includegraphics[width=1\linewidth]{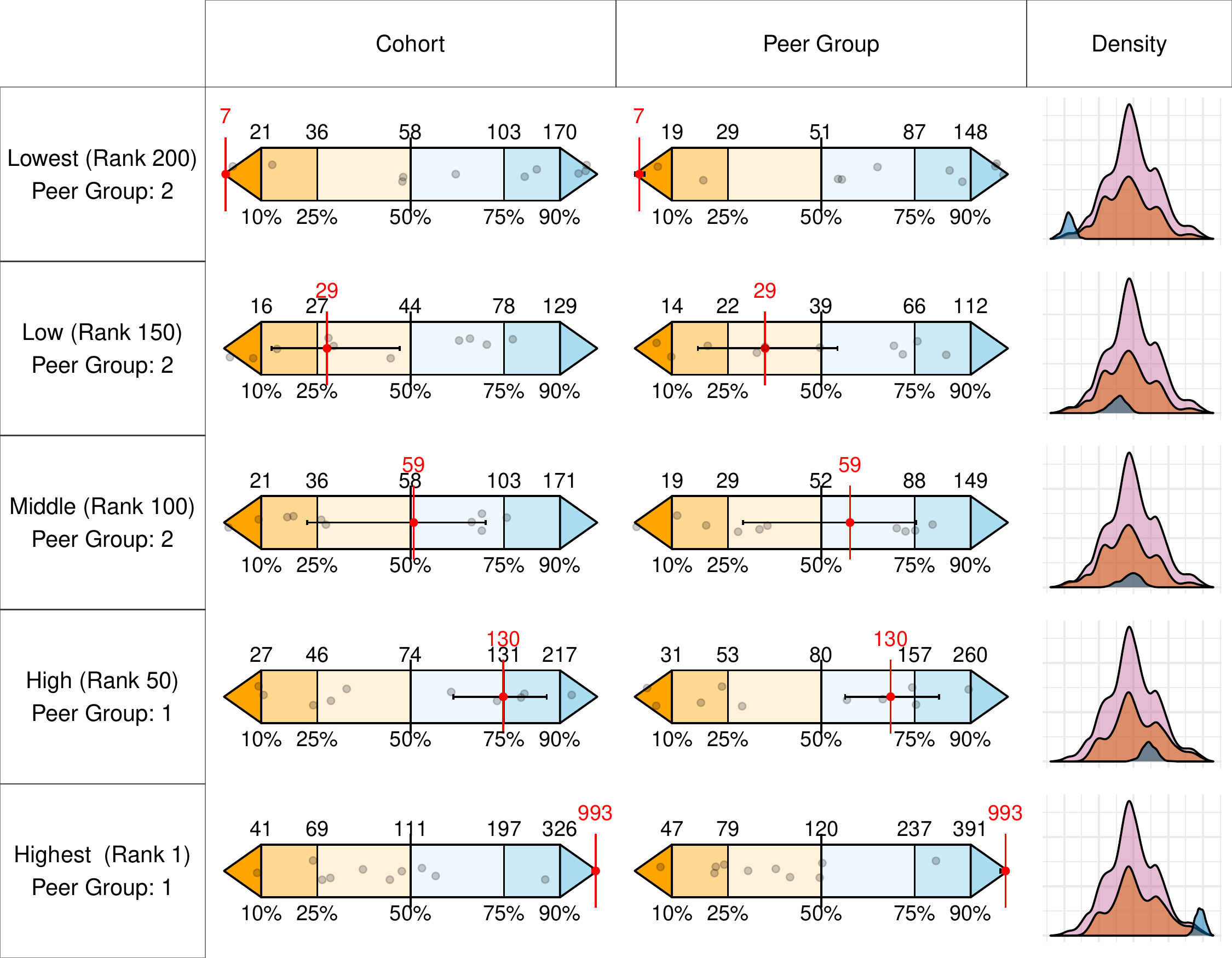}
	\caption{Spirit level plots showing the comparative measure of organisations with uncertainty. The left column is for measure relative to the entire cohort, and right column is for measure relative to the organisation's peer group. The x-axis scale is the percentiles of the bootstrap residual distribution. Organisations were chosen representing the smallest, largest, and quartiles by mean residual value. The final column, which could be omitted if desired, shows the full cohort residual distribution (purple), the peer-group residual distribution (orange), and the organisation's own residual distribution (blue).}
	\label{fig:combinedadjperfplot}
\end{figure}

\section{Discussion}

In this paper we fitted a random forest model as the adjustment model for a comparative measure for a data-set with complex contextual factor effects, and incorporated the effects of peer grouping.
This model choice, which was informed by out-of-sample predictive accuracy, was shown to outperform the linear regression method in both accuracy, and in reducing bias in estimation of each organisation's adjusted measure.
We developed a benchmark using bootstrapping relative to which organisations' measure values can be evaluated, and a procedure for approximating rank uncertainty.
Importantly, the benchmark that an organisation is compared against can be cohort-level, or specific to their peer-group, to ensure comparison are valid. We used a novel visualisation as means of communicating the random forest model and the adjusted measures to technical and non-technical audiences.

Although the random forest was used for this work, most of the analysis presented could be described as model-agnostic.
The bootstrapping for residual distributions, normal approximation for ranking, as well as the spirit-level and percentile plots could all be used with other statistical learning regression models.
The importance measures used here are specific to random forests, however, variable importance has been studied for Generalised Additive Models~\citep{Goetz2011}, linear regression~\citep{Grmping2015}, and gradient boosting machines~\citep{Friedman2001}.
We found in this case that the random forest out-performed linear regression, but this is likely to be a function of the data.
In general, we recommend using an out-of-sample prediction accuracy measure to determine the best model for benchmarking.

The random forest in this case study performed well because of nonlinear relationships between the response and the covariates.
The step-wise relationship found for covariate 8 close to its minimum value is easily modelled by the random forest, since the binary cuts of regression trees approximate step function relationships well.
The linear regression by contrast failed to approximate this relationship well as well as non-monotonic relationships seen in other variables.
In addition to superior accuracy, the random forest's inbuilt importance measure is useful for identifying which variables are key to predicting the comparative measure, and thus important for adjusting.
A second advantage of the random forest is the out-of-bag prediction, which is useful for preventing overfitting in predictions. However, for bootstrap approximation of the predictive distribution is was critical to include only out-of-resample replicates to prevent a strong negative bias in residual uncertainty estimates.
This is important because to ensure fair comparisons, adjusted measures must be independent of contextual factors and be presented with an accurate estimate of uncertainty to avoid overconfidence in ranking or comparisons to a benchmark.

For parametric models such as linear regression, the value of the standard error is easily attributable based on proximity of the given observation to the data mean, however this is not the case for random forests.
The uncertainty in rank confidence intervals was very well explained by the mean and standard error of the residuals, which is to be expected.
However, there was more difficulty in explaining the standard error of residuals themselves as the proximity of an observation to other observations only partially explained it.
It is likely that regions where there is a strong effect of covariates have greater uncertainty than at locations where the mean response is relatively constant.
Therefore, Euclidean distance may not be a sophisticated enough measure to describe proximity, and further work may reveal a measure which can better reflect the available information in the data for prediction at a given point, and thus the residual uncertainty.

A limitation of this work is that the bootstrapping scheme cannot directly recover full rankings.
Our solution, the use of a normal approximation to the joint residual distribution, is only correct to mean and standard deviation, and it was evident there was small deviations in correlation between residuals.
A more sophisticated approach could use available three-way and higher order correlations to achieve a closer approximation to the bootstrap distribution.
Another approach could leverage the fact that for each bootstrap replicate, it is possible to obtain a partial ranking based on the non-missing bootstrap residuals, so ranking aggregation procedures may be useful to establish a full representative ranking (For an example see \citet{Ailon2008}).

The bootstrap provided a means of approximating the predictive distribution of the residuals given model uncertainty.
A key assumption of the use of it as a benchmark is that the error distribution is homogeneous over all organisations.
This assumption is partially relaxed by making peer-group-specific distributions, however it does not control for sub-peer-group changes in the error distribution.
One option might be to use only the closest neighbouring observations to form a predictive distribution for a given observation. However, determining how many neighbours are used requires balancing a trade-off between the number of observations used and how locally-specific the predictive distribution is.

We found that there were only small differences in the bootstrap residual distributions between peer-groups, indicating a similar spread of comparative measures for organisations within them.
This was further seen in the spirit-level plots, where the relative adjusted measure percentiles for organisations did not change markedly. This is however not an indication that comparisons such as difference calculation or ranking can be done between organisations across peer-groups in general.
As in the industry organisation economics literature for strategic groups, disparate organisations may employ the same strategy yet yield different results in terms of performance. It is therefore imperative that comparisons be made only within peer-groups of similar organisations.


The critical challenges in benchmarking are developing appropriate and meaningful metrics for comparing organisations from data, and then interpreting those metrics in context so as to ensure fair and valuable comparisons.
This work has produced tools for the second challenge, using a principled, statistical learning approach.
While it is tempting to account for contextual factors using linear regression, we have shown that the random forest provides a more comprehensive adjustment in real data with complex effects.
We have shown that the random forest adjustment model can be explored and understood by computing relative variable importance and partial dependence plots, and that outputs such as ranking and comparison with a benchmark can be computed with bootstrap-approximated uncertainty.
Finally, we have developed visualisations which can be interpreted with minimal domain knowledge, expressing both the adjusted measure relative to an intuitive benchmark, ensuring all stakeholders are able to use this information for organisational learning.

\bibliographystyle{imsart-nameyear} 

\bibliography{benchmarking-paper-references}

\appendix

\begin{appendix}
\section*{Tuning via Cross-Validation}
\begin{figure}[h]
	\includegraphics[width = 1\linewidth]{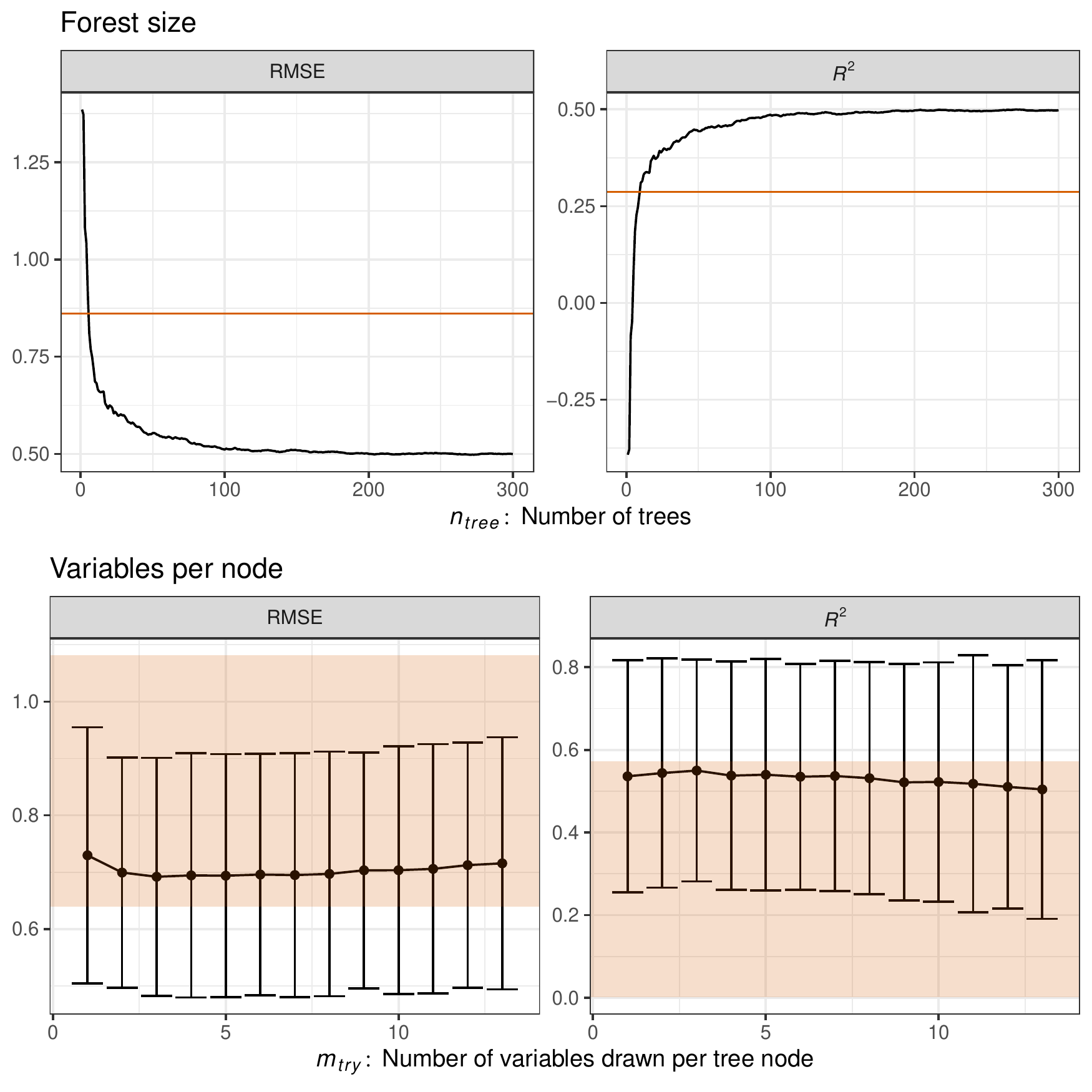}
	\caption{(top) Out-of-Bag measures of prediction accuracy as a function of the number of trees in the forest. Both measures indicate 100 trees is sufficient to achieve near-optimal accuracy. (bottom) Measures of generalised accuracy calculated using repeated k-fold cross-validation. Accuracy measures Root Mean Squared Error (RMSE) and Pearson's $R^2$ are calculated for values of $m_{try}$ from 1 to 14. A greater value of $m_{try}$ increases the complexity of the forest and the variance, however small values can cause bias. Standard errors of these values are depicted using the error bars, which span one standard error above and below the estimate.}
\end{figure}
\end{appendix}
%
%


\section*{Acknowledgements}

The authors would like to acknowledge the support of the ARC Centre of Excellence for Statistical and Mathematical Frontiers.

KM and DK are supported by an ARC Laureate Fellowship (FL150100150).

The authors would like to acknowledge the agency which contributed financial support towards this research.





\end{document}